\documentclass[twocolumn,trackchanges]{aastex631}

\usepackage[T1]{fontenc} 

\begin{document}

\title{Exploring the properties of photosphere and emission lines for tidal disruption events based on the global solution of slim disk and winds}

\author[0009-0001-7472-615X]{Yuehua Zhang}
\affiliation{Department of Astronomy, School of Physics, Huazhong University of Science and Technology, Luoyu Road 1037, Wuhan, China}

\author[0000-0003-4773-4987]{Qingwen Wu$^*$}
\affiliation{Department of Astronomy, School of Physics, Huazhong University of Science and Technology, Luoyu Road 1037, Wuhan, China}

\author[0000-0002-2581-8154]{Jiancheng Wu}
\affiliation{Department of Astronomy, School of Physics, Huazhong University of Science and Technology, Luoyu Road 1037, Wuhan, China}

\author[0000-0002-2355-3498]{Xinwu Cao}
\affiliation{Institute for Astronomy, School of Physics, Zhejiang University, 866 Yuhangtang Road, Hangzhou 310058, People’s Republic of China}

\author[0000-0003-3440-1526]{Weihua Lei}
\affiliation{Department of Astronomy, School of Physics, Huazhong University of Science and Technology, Luoyu Road 1037, Wuhan, China}

\begin{abstract} \label{sec:abs}
The theoretical debris supply rate from a tidal disruption of stars can exceed about one hundred times of the Eddington accretion rate for a $10^{6-7}M_{\odot}$ supermassive black hole (SMBH). It is believed that a strong wind will be launched from the disk surface due to the radiation pressure in the case of super-Eddington accretion, which may be one of the mechanisms for formation of the envelope as observed in tidal disruption events (TDEs). In this work, we explore the evolution of the envelope that formed from the optical thick winds by solving the global solution of the slim-disk model. Our model can roughly reproduce the typical temperature, luminosity and size of the photosphere for TDEs. Based on \texttt{CLOUDY} modeling, we find that, if only considering the radiation-driven disk wind, the emission line luminosities are normally much lower than the typical observational results due to the limited atmosphere mass outside the envelope. We propose that the ejection of the outflow from the self-collision of the stellar debris during the circularization may provide enough matter outside the disk-wind photosphere. Our calculated spectra can roughly reproduce the main properties of several typical emission lines (e.g., $\rm H\alpha$, $\rm H\beta$ and \ion{He}{2}), which was applied well to a TDE candidate AT2018dyb.

\end{abstract}


\section{Introduction} \label{sec:intro}
When a star passes too close to a supermassive black hole (SMBH), the tidal force of the black hole (BH) may exceed the star's self-gravity, leading to the disruption of the star \citep[so-called tidal disruption event, TDE, e.g.,][]{rees1988tidal,phinney1989manifestations}. After disruption, roughly half of the material will escape form the SMBH, while the remaining will fall back and eventually be accreted by the central SMBH. \citet{rees1988tidal} suggested that the evolution of TDE luminosity roughly follows $L(t)\propto t^{-5/2}$, which was later corrected to $L(t)\propto t^{-5/3}$ by \cite{phinney1989manifestations}. The multi-waveband flares of TDE usually evolve on time scales of a few months and play an important role in exploring quiescent SMBHs with masses $M_\mathrm{BH} \lesssim 10^8 M_{\odot}$ \citep{saxton2020x, van2020optical, gezari2021tidal}. TDEs were theoretically predicted almost fifty years ago \citep{hills1975possible}, however, detections of such exotic events are much later, which are first detected in the X-ray wavebands \citep{komossa1999giant, grupe1999rx}, and followed by the ultra-violet (UV) \citep{gezari2006ultraviolet} and optical wavebands \citep{gezari2012ultraviolet}. The number of candidate TDEs is rapidly growing in the era of wide-field optical transient surveys, including ASAS-SN \citep[e.g.,][]{holoien2014asassn, holoien2016six, holoien2016asassn, holoien2019discovery, hinkle2021discovery}, ZTF \citep[e.g.,][]{van2019first, van2021seventeen, hammerstein2022final}, SDSS \citep[e.g.,][]{abazajian2009seventh, van2011optical}, iPTF \citep[e.g.,][]{hung2017revisiting, blagorodnova2017iptf16fnl, blagorodnova2019broad} and Pan-STARRS \citep[e.g.,][]{gezari2012ultraviolet, chornock2013ultraviolet, holoien2019ps18kh}. With more and more observations, it is found that TDEs are possibly overrepresented in post-starburst or "green valley" galaxies, which suggests that the dynamics of stars in the nuclear environments is strongly evolved with the galaxy evolution \citep{yao2023tidal, hammerstein2023integral, wang2023explanation}.

For BHs with masses around $\sim 10^6-10^7 M_{\odot}$, the fallback rate of TDEs can exceed the Eddington accretion rate for a couple of months. The classical scenario of TDE emission comes from disk radiation, where the accretion disk is formed after the circularization of the bound debris as it falls back to the central SMBH \citep{rees1988tidal, phinney1989manifestations}. If this is the case, most of the emission should be radiated in the soft X-ray waveband. However, most TDEs are detected in the optical waveband without strong X-ray emission, where the continuum temperature derived from optical/UV bands (e.g., several tens of thousands Kelvin) is usually an order of magnitude lower than that of X-ray TDEs \citep{van2011optical}. The optical emission is widely believed to originate in the envelope surrounding the SMBH, which absorbs the X-ray emission produced by the accretion disk and reradiates at the optical waveband \citep[e.g.,][]{loeb1997optical, strubbe2009optical, roth2016x, piro2020wind}. Such an optical envelope can be formed through debris-collision-induced outflow \citep[e.g.,][]{guillochon2014possible, coughlin2014hyperaccretion, shiokawa2015general, guillochon2015dark, cao2018failed, lu2020self}, or from disk wind \citep[e.g.,][]{lodato2011multiband, uno2020application, mageshwaran2023probing}. The optical TDEs with possible rebrightening in the UV light curve was explained by the collision of debris and later the formation of the accretion disk \citep{guo2023evidence, chen2021light}. Apart from the X-ray TDEs, some initially optically strong TDEs brighten in X-rays several tens of days after the peak of optical emission \citep[e.g.,][]{gezari2017x, holoien2018unusual, wevers2019evidence, hinkle2021discovery, liu2022uv}. A small fraction of TDEs are also discovered in the mid-infrared \citep[e.g.,][]{mattila2018dust, jiang2021infrared} and radio bands \citep[e.g.][]{van2014measurement, arcavi2014continuum}, which are possibly from the torus and jet/outflow, respectively. \citet{dai2018unified} proposed a unified model for TDEs, which includes an accretion disk, an envelope triggered by optically thick winds, and a possible jet. This unified model explained the optical and X-ray emission based on viewing-angle effects, where the soft X-ray and optical emission are viewed from small and large viewing angles, respectively.

Early spectroscopic observations suggested that TDEs show mainly a range of spectral properties with broad He and/or H lines \citep{arcavi2014continuum}. However, recent observations on more TDEs revealed a greater diversity than previously thought\citep[e.g.,][]{short2020tidal, hung2020double, charalampopoulos2022detailed}. Nitrogen and oxygen lines are also discovered \citep[e.g.,][]{blagorodnova2019broad, leloudas2019spectral, onori2019optical}, which are attributed to the Bowen fluorescence mechanism \citep{bowen1934excitation, bowen1935spectrum}. Based on spectropic studies on a sample of TDEs, \citet{hammerstein2022final} divided the TDE spectrum into four categories: TDE-H, TDE-He, TDE-H+He, and TDE-featureless. The line profile is normally very broad ($\sim 10^4 \rm km \ s^{-1}$) and shows a large diversity as well as different line profiles, with some TDEs even showing double peak profiles as those observed in active galactic nuclei \citep[AGN, e.g.,][]{cao2018large, short2020tidal, hung2020double}. The properties of emission lines can shed light on the ionization source, dynamics, and optical depth of the emitting atmospheres. For the observed Bowen lines, there must be a source of X-ray/far-ultraviolet (FUV) photons, which is needed to trigger a cascade of transitions and eventually lead to the high-ionization Nitrogen and Oxygen lines \citep{leloudas2019spectral}. The flux ratio of emission lines also provides insight into the optical depth of emitting atmospheres. \citet{roth2016x} suggested that the strength of lines in TDEs is set by the wavelength-dependent optical depths, and proposed that the helium photosphere should lie deeper than the hydrogen one in a stratified atmosphere. The gas in inner photosphere and outer atmosphere may not be virialized, and the line width cannot be used to estimate the BH mass. \citet{saxton2018spectral} proposed an alternative light-echo photoionization model for the TDE spectrum, where the initial flash of a hot blackbody excites the outer atmosphere, and then is seen superimposed on continuum from a later, expanded, cooled stage of the luminous source. This is reasonable if the emitting atmosphere stays outside of the photosphere. Reverbration mapping on TDE sources can test this light-echo scenario. On the physical mechanism of line width, \citet{roth2018sets} proposed that line width may be set by the optical depth of electron scattering rather than by gas kinematics. Further studies on the evolution of line profiles can shed light on the properties of emitting atmospheres in TDEs.

The observed peak luminosity of most TDEs is sub-Eddington, which is inconsistent with the estimate from the fallback rate (exceeds the Eddington accretion rate by two orders of magnitude at the peak). One possibility is that the intrinsic spectral energy distribution (SED) has a much broader shape than the single temperature spectrum adopted in modeling the TDE spectrum \citep{thomsen2022dynamical}. The second possibility is that the photons lose energy before escaping from the outflowing gas and convert part of the energy into the kinetic energy of the outflow. Considering the possible super-Eddington accretion in TDEs, the slim disk should be adopted to explore the possible physical properties of TDEs\citep[e.g.,][]{thomsen2022dynamical}. It should be noted that strong winds will naturally be launched due to the radiation pressure if the accretion rate is larger than several times of Eddington rate \citep{cao2015limits, feng2019global, cao2022supercritical}. In this paper, we explore the evolution of the photosphere based on the global solution of slim disk, where the envelope is formed through the disk wind. We further calculate the intensity of the emission lines from the atmosphere that stay outside of the photosphere based on the photoiniozation code \texttt{CLOUDY}. 

In section \ref{sec:2}, we present the models of disk wind, envelope and atmosphere. In section \ref{sec:3}, we present our main results. The conclusion and discussion are given in section \ref{sec:4}.

\section{Models} \label{sec:2}
\begin{figure}[ht!]
\plotone{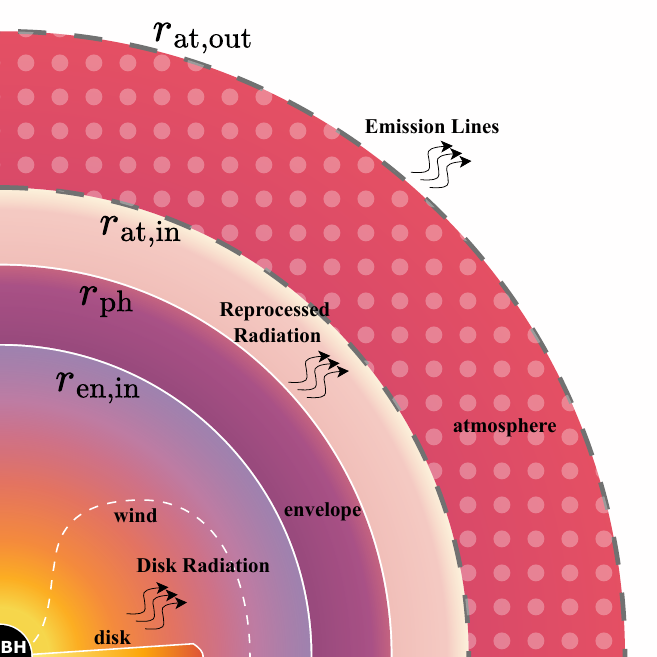}
\caption{Schematic of our model. The disk wind is formed and accelerated by the radiation pressure, and the part of optically thick winds will form the envelope, where the envelope will reprocesses the disk emission. The emission lines are produced at the atmosphere outside of the envelope, which can be contributed from the disk wind and/or the collision shock of stellar debris before the disk formation.
}
\label{fig:1}
\end{figure}

The formation, geometry, and evolution of the envelope should be well considered in order to explore the properties of the photosphere and emission lines. However, all of these processes have been poorly studied. In this work, we consider the outflowing wind model based on an analytical slim disk model with global solution covering the super-to-sub-Eddington regimes. We find that the predicted line luminosities from the pure disk wind outside the optically thick envelope are much lower than the observational value (as described in more detail below). Therefore, the original sourrounding interstellar medium or the material thrown out from the stream-stream collisions may also contribute to the atmosphere outside of the photosphere. More importantly, the density of the atmosphere is most possibly not following a simple power-law profile as many works adopted, since that the fallback rate is strongly decreases with time. We describe our models in more detail as follows. The main structure of our model is shown schematically in Figure \ref{fig:1}.

\subsection{Slim disk with radiation-driven wind}
After the disruption of the star, about half of the stellar debris will be captured by the central SMBH, and the fallback rate often approximately follows 
the $t^{-5/3}$ scaling \citep[e.g.,][]{rees1988tidal, phinney1989manifestations},
\begin{equation}  \label{eq:1}
\dot M_\mathrm{fb} \sim \dot M_\mathrm{max}(\frac{t}{t_\mathrm{mb}})^{-\frac 53}, 
\end{equation}
where $\dot M_\mathrm{max}$ is the maximum fallback rate and $t_\mathrm{mb}$ is a parameter related to the circularized orbital period. $\dot M_\mathrm{fb}$ can exceed the Eddington rate ($\dot{M}_{\rm Edd}=1.44\times 10^{18} M_\mathrm{BH}/M_\odot$ $\rm g$ $\rm {s}^{-1}$) by two orders of magnitude at peak.

In the super-Eddington accretion phase, the disk temperature will increase, and the disk will become thicker. Considering the radiation pressure and possible advection of the magnetic field with the accreting material, the disk winds can be launched easily in the regime of super-Eddington accretion \citep[e.g.,][]{bisnovatyi1977disk, cao2015limits}. Furthermore, the photon trapping effect will play an important role in the geometrically thick slim disk, and a fraction of photons cannot escape the disk and will eventually be advected into the SMBH \citep{begelman1978black}. Therefore, the radiative efficiency of the slim disk is lower than that of standard accretion disk, and the disk luminosity is roughly saturated at very high accretion rates \citep{watarai2006new, cao2015limits}. \citet{feng2019global} explored the global solution of the slim disk, and found that the mass accretion rate decreases with the decreasing radius in the disk with winds and the rate of gas swallowed by the BH is always limited to $\sim1.7-1.9 \dot{M}_{\rm Edd}$ even if the mass accretion rate at the outer edge of the disk is very high. Strong, dense optical winds can form the envelope above the slim disk. 

To calculate the possible envelope as found in the TDEs from the disk-wind model, we calculate the global structure of the slim disk following \citet{feng2019global}, which are simply summarized as follows \citep[see][for the details]{kato1998black, gu2007note, feng2019global}. We consider a steady axisymmetric accretion disk, which is described by the momentum equation, the energy equation, the continuity equation, and the equation of state. Taking into account the hydrostatic equilibrium with the vertical radiation flow, the wind will be launched when the radiative flux $F^{-}_r>f_{\rm max}=(2\sqrt{3/9})(c/\kappa)GM_{\rm BH}/r^2$ \citep{cao2015limits}, where $c$ is the light speed and $\kappa$ is the opacity of the accretion flow. It should be noted that disk wind can also appear in the sub-Eddington state as constrained from both BH X-ray binaries and AGNs, which may be driven by some other physical mechanisms (e.g., magnetic field and/or line driven, etc.)\citep[e.g.,][]{proga2000winds, nomura2016radiation}. To include this effect and allow wind production in the sub-Eddington regime, we describe the accretion rate as $\dot M_\mathrm{d}\propto r^s$ in our global solution of the slim disk. It should be noted that more gas at the disk surface will be further driven into outflows by radiation force in the super-Eddington case when the disk thickness surpasses the critical value. The global structure of the slim disk and the outflow rate can be derived with parameters of viscosity ($\alpha$), BH mass ($M_{\rm BH}$) and accretion rate ($\dot{M}$) in the outer radius. In this work, we simply adopt $\alpha=0.1$, $\dot{M}_{\rm max}=100\dot{M}_{\rm Edd}$ at $r_{\rm out}=1000r_{\rm s}$ ($r_{\rm s}=2GM_{\rm BH}/c^2$ is the Schwarzschild radius).

\subsection{Wind Envelope}

Following previous work\citep{uno2020application, mageshwaran2023probing}, we assume that there is an optically thick envelope outside the accretion disk, which reprocesses the radiation from the underlying disk. In the above slim-disk model, the outflow rate can be naturally derived at a given accretion rate. The exact geometry and evolution of the wind in the TDEs are still unclear. We assume that the velocity of the wind in the envelope is $v_{\rm w}$, and the density profile should be
\begin{equation}  \label{eq:2}
\rho_\mathrm{en}(r)=\frac{\dot M_\mathrm{w}}{4\pi r^2 v_\mathrm{w}},
\end{equation}
where $\dot{M}_{\rm w}$ is outflow rate. The wind will be accelerated, and the kinetic energy of the wind will eventually be in equilibrium with the internal energy dominated by radiation at a certain radius. We define this equilibrium radius as the inner radius of the envelope $r_\mathrm{en,in}$. When $r_\mathrm{en,in}$ is given, we can derive the temperature $T_\mathrm{en,in}$ at $r_\mathrm{en,in}$ by $1/2\rho(r_{\rm en,in}) v^2_{\rm w}\sim aT_\mathrm{en,in}^4$.

Assuming the outer radius of the envelope satisfies $r_\mathrm{en,out}\gg r_\mathrm{en,in}$, the scattering optical depth is given by 
\begin{eqnarray}  \label{eq:3}
    \tau(r)&=&\int_{r}^{\infty}\kappa_\mathrm{es} \rho_\mathrm{en}(r)\mathrm{d}r  
\end{eqnarray}
where $\kappa_{\mathrm {es}}$ is the electron scattering opacity.

Near the inner radius of the envelope ($r_{\rm en,in}$), the density is high and the photon diffusion time may be longer than the dynamical time, where the photons are trapped and coupled with the matter. The photon-trapping radius $r_\mathrm{en,tr}$ of the photosphere is
\begin{equation}  \label{eq:4}
\frac{r_\mathrm{en,tr}-r_\mathrm{en,in}}{v_\mathrm{w}}=\frac{r_\mathrm{en,tr}\tau(r_\mathrm{en,tr})}{c}.
\end{equation}

Within the photon-trapping radius, $r_\mathrm{en,in}<r<r_\mathrm{en,tr}$, the gas evolved roughly adiabatically. The temperature evolves as $T\propto r^{-2/3}$. Above the photon-trapping radius, $r>r_\mathrm{en,tr}$, the photon will diffuse through the wind, the luminosity is 
\begin{equation}  
L=-\frac{4\pi r^2ac}{3\kappa_\mathrm{es}\rho_\mathrm{en}(r)}\frac{\partial T^4}{\partial r}, 
\end{equation}
where the luminosity is nearly constant, and the temperature evolves as $T\propto r^{-3/4}$ .

In the diffusive region, the absorption of photons by the electrons can also influence the opacity, and the total opacity should be given by $\kappa_\mathrm{eff}=\sqrt{3\kappa_\mathrm{es}\kappa_\mathrm{a}}$, where $\kappa_\mathrm{a}=\kappa_0 \rho_\mathrm{en} T^{-7/2} \mathrm{cm}^2 \mathrm{g}^{-1}$\citep[see][for the details]{uno2020application}.

Considering $r_\mathrm{en,out}\gg r_\mathrm{en,in}$, the color radius can be derived from
\begin{equation}  \label{eq:5}
\int_{r_\mathrm{en,cl}}^\infty\kappa_\mathrm{eff}(r) \rho_\mathrm{en}(r) \mathrm{d}r=1.
\end{equation}

The photospheric radius is therefore given by
\begin{equation}  \label{eq:6}
r_\mathrm{ph}=\mathrm{max}\{r_\mathrm{en,tr},r_\mathrm{en,cl}\},
\end{equation}

and the effective photospheric temperature is
\begin{equation}  \label{eq:7}
T_\mathrm{ph}=(\frac{L}{4\pi r_\mathrm{ph}^2\sigma_\mathrm{sb}})^{\frac 14},
\end{equation}
where $\sigma_\mathrm{sb}$ is Stefan-Boltzmann constant.

With the above equations, we can get the photospheric radius $r_\mathrm{ph}$, photospheric temperature $T_\mathrm{ph}$ and blackbody radiation luminosity $L_\mathrm{ph}$ if the wind velocity is known. In our model, a fraction of the disk radiation energy is converted into the photosphere radiation, and the other part is converted into the kinetic energy of the wind, where $L_\mathrm{ph}+\frac 12 \dot M_\mathrm{w} v_\mathrm{w}^2=L_\mathrm{disk}$ is assumed.

\subsection{Atmosphere and Emission-line calculation}
 The photoionization of the atmosphere by the inner photosphere will produce the emission lines. The origin of the atmosphere is not clear, which may come from the preexisting circumnuclear medium surrounding the SMBH, outflowing gas from debris collision, and/or the disk wind during accretion phase\citep[e.g.,][]{jiang2016prompt, skadowski2016magnetohydrodynamical, lu2020self}. In our calculations, we find that the mass of the envelope contributed from the pure disk winds is too low (normally less than several percent of the solar mass) to produce the observed peak line luminosities. Therefore, we assume that the initial debris collision also contribute to the outer atmosphere since that the possible preexisting circumnuclear medium is even more unclear \citep{Zhou2024}.

\begin{figure}[ht!]
\includegraphics[width=\columnwidth]{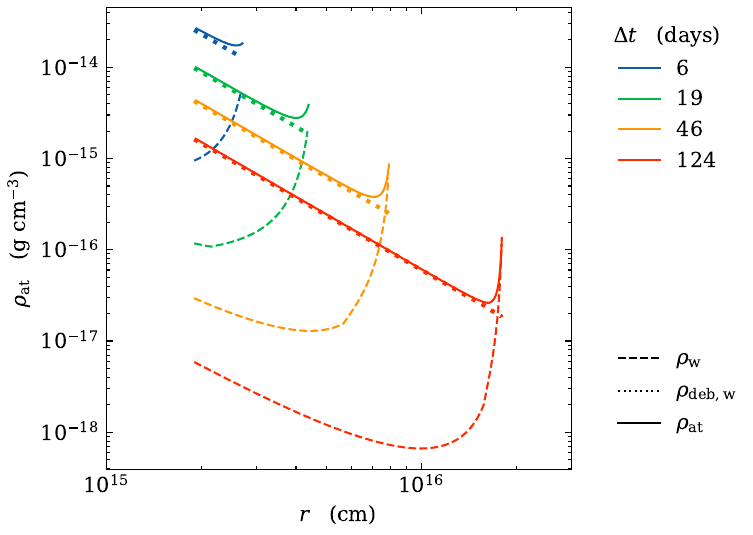}
\caption{The densities in the atmosphere $\rho_\mathrm{at}(r)$ as functions of radius, where $\Delta t$ can simply determine the outer radius of the atmosphere (e.g., $r_{\rm at,out}=r_{\rm at,in}+v_{\rm at}\times \Delta t$). The dashed and dotted lines represent the contributions of wind and debris respectively, and the solid lines represent the overall density distribution. The parameters we adopted are $\rho_\mathrm{i0}=1\times 10^{-13} \mathrm{g}$ $\mathrm{cm}^{-3}$, $p_\mathrm{i}=1$ , $v_{\rm at}=0.05c$ and $r_\mathrm{at,in}=2\times 10^{15} \mathrm{cm}$ as an example. }
\label{fig:2}
\end{figure}

 Considering the fast evolution of the fallback rate, the density distribution of outflowing wind should not follow a simple power-law distribution, i.e., $\rho (r)\propto r^{-p}$ with $p=2$ \citep[for a steady state wind, e.g.,][]{uno2020application} is normally adopted. In this work, we assume that the atmosphere outside of the photosphere is contributed by both debris collisions during circularization and disk wind in the accretion phase. The density of the atmosphere at the inner radius $r_\mathrm{at,in}$ follows
\begin{equation}  \label{eq:8}
\rho_\mathrm{at,in}(t)=\rho_\mathrm{deb,w}(t)+\rho_\mathrm{w}(t),
\end{equation}
where $\rho_\mathrm{deb,w}(t)$ is the density contributed by the outflow from the debris collision and $\rho_\mathrm{w}(t)$ is contributed by winds from accretion disk. The collision of debris will become weaker with decreasing fallback rate. In this work, we simply assume that
\begin{equation}  \label{eq:9}
\rho_\mathrm{deb,w}(t)=\rho_\mathrm{i0}\left(\frac{t}{t_\mathrm{mb}}\right)^{-p_\mathrm{i}},
\end{equation}
where $\rho_\mathrm{i0}$ is initial density at $r_{\rm at,in}$ at given time. The mass of the atmosphere is determined mainly by the density of the mass and the distribution of the density, which will regulate the intensity of the line emission.

Assuming that the gas in atmosphere also expands out with a velocity $v_\mathrm{at}$, the density of expanding matter follows $\rho\propto r^{-2}$ to ensure mass conservation. Considering evolution of $\rho_\mathrm{at,in}(t)$ with $\dot{M}(t)$, the real $r$ denpendent density distribution of the atmosphere $\rho_\mathrm{at}(r,t)$ will not follow a simple power law at given time $t$, where the higher density at earlier time will move to larger scale. For example, we present the density profile of the outer atmosphere in Figure \ref{fig:2}, where $M_\mathrm{BH}=5\times 10^6 M_\odot$, $s=0.15$, $v_\mathrm{at}=0.05c$, $\rho_\mathrm{i0}=1\times 10^{-13} \mathrm{g}$ $\mathrm{cm}^{-3}$, $p_\mathrm{i}=1$ and $r_\mathrm{at,in}=2\times 10^{15} \mathrm{cm}$ are adopted. The dashed and dotted lines represent the contribution from the debris collision and disk wind respectively, while the solid lines are the total contribution. In our atmosphere model, the outflow from the debris collision will be dominant, and the disk wind only dominate in large scale at sometime later.

With ionization luminosity calculated from the photosphere and the atmosphere as shown above, we adopt the photoniozation code \texttt{CLOUDY} (c21.00) to model the optical emission lines for TDEs, which consider the detailed microphysics to simulate the physical conditions of nonequilibrium gas clouds exposed to an external radiation field. The thermal, statistical and chemical equilibrium equations can be solved self-consistently with initial conditions (e.g., density distribution, chemical composition, radiation field), where this \texttt{CLOUDY} model has been used in various astrophysical environments \citep[see][for the details]{ferland20132013,ferland20172017}. For the case of TDEs, \citet{saxton2018spectral} found that standard AGN-like irradiation cannot explain both the line ratios and the individual equivalent widths of TDEs simultaneously. \citet{sheng2021evidence} also tried to constrain the C/N ratio for a possible TDE candidate of GSN 069 based on an AGN-like irradiation, where the line-emitting gas is assumed to come from the interior of a star. However, GSN 069 is an X-ray TDE candidate, which maybe much different from the optical TDEs, where their radiation fields are different. \cite{mageshwaran2023probing} explored the atmospheric properties based on a disk-wind model and further constrain TDE line properties based on CLOUDY modelling. Based on a simlified disk-wind model, a very large disrupted star $\sim 6 M_{\odot}$ is needed to reproduce the observed line luminosities.

In this work, we aim to further explore the properties of broad lines in TDEs based on envelope as formed from wind of slim disk and a spherical atmosphere that contributed from the disk wind and/or the collision shock of stellar debris before the disk formation. In the spectrum calculations, we adopt the time-independent model, which is much faster than the time-dependent one. However, the accretion rate of black holes and the density (or mass) of the atmosphere are time-dependent, which can mimic the possible evolution of TDEs. This calculation method was also widely used in earlier work because it can save a lot of computing time \citep[e.g.,][]{pandey2022photoionization, mageshwaran2023probing}. This simplified calculation is reasonable if the formation and recombination timescales are much smaller than the age of the cloud \citep{mageshwaran2023probing}. After assuming the evolution of accretion rate, we calculate the radiation filed from the photospere at each epoch, and then calculate the line luminosities using CLOUDY model with inner/outer boundaries and density distribution of atmosphere(e.g., as shown in Figure \ref{fig:2}). Due to the unknown time interval between the debris collision and disk formation, we set the inner radius of atmosphere ($r_\mathrm{at,in}$) as a free parameter. For the outer boundary of the atmosphere, we self-consistently calculated from the expanding velocity and the time interval from launching (e.g., $r_{\rm at,out}=r_{\rm at,in}+v_{\rm at}\times \Delta t$). The solar metallicity is adopted, even though the metallicity of different disrupted stars may be different.

\section{Results} \label{sec:3}

\subsection{Wind envelope and \texttt{CLOUDY} results}

Based on global solutions of the slim disk, the strong winds are driven from the disk surface when $\dot{m}>\dot{m}_{\rm c}\sim1.7-1.9$. The envelope surrounding the disk will be formed if the winds are optically thick. We assume a spherical envelope expanding with velocity $v_\mathrm{w}$, where the inner radius of the envelope $r_{\rm en,in}$ is assumed to be the equilibrium place of internal energy and kinetic energy. The inner radius of the envelope $r_{\rm en,in}$ is normally adopted as a free parameter and can be constrained from fitting the observational data \citep[e.g.,][]{uno2020application, mageshwaran2023probing}. To show our theoretical results, we simply assume that the inner radius expands up to $r_\mathrm{l0}$ with constant speed at the early stage (at this stage $r_{\rm en,in}$ is proportional to $t$) and shrinks with a power-law form $r_{\rm en,in}=r_\mathrm{l0}t^{-p_{rl}}$ at the later stage, respectively, which can roughly reproduce the observed evolution of photosphere size \citep[e.g., increase and then decay,][]{charalampopoulos2022detailed}. 

\begin{figure*}[ht!]
\plotone{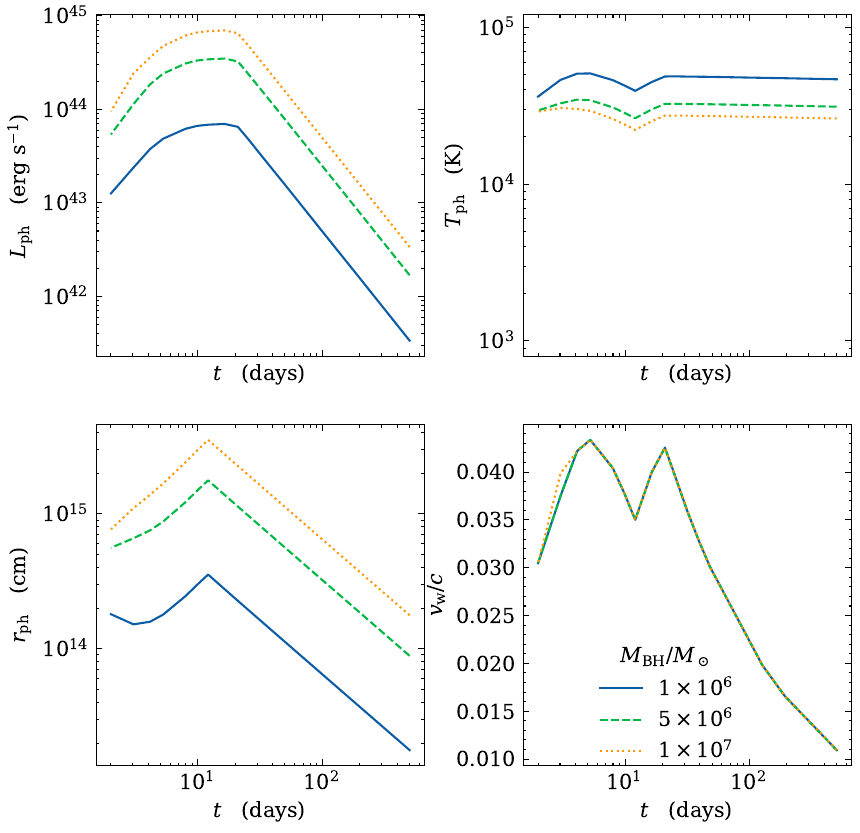}
\caption{An example of the evolution for photosphere lumnosities $L_\mathrm{ph}$ (top-left panel), radii $r_\mathrm{ph}$ (bottom-left panel), temperatures $T_\mathrm{ph}$ (top-right panel) and wind velocities $v_\mathrm{w}$ (botttom-right panel) with time. In this example, the parameters of $s=0.1$, $r_{\mathrm{l}0}/r_\mathrm{s}=1200$ and $p_\mathrm{rl}=0.8$ are adopted.}
\label{fig:3}
\end{figure*}

In Figure \ref{fig:3}, we present an example for the evolution of luminosity, size, and temperature of the photosphere for typical parameters of TDEs. In the disk-wind model, the disk luminosity will be roughly saturated in super Eddington accretion phase. Therefore, the luminosity of the reprocessed photosphere just approaches $\sim$ Eddington luminosity, and a fraction of the disk energy is converted into the kinetic energy of outflow. In the super-Eddington phase, strong winds will form and the envelope will expand. As a result, both the luminosity and the size of the photosphere will increase. The winds will become weaker as the accretion rate decreases, and the luminosity and size of the photosphere will decrease. To maintain energy conservation, the velocity of the winds also increases in the initial stage and then decays. During this evolution, the temperature of the photosphere is roughly unchanged. It should be noted that the zigzags in the top-right and bottom-right panels of Figure \ref{fig:3} at $t\sim 10 \rm days$ are caused by the unsmooth connection of the power-law winds and the slim-disk wind.

\begin{figure*}[ht!]
\plotone{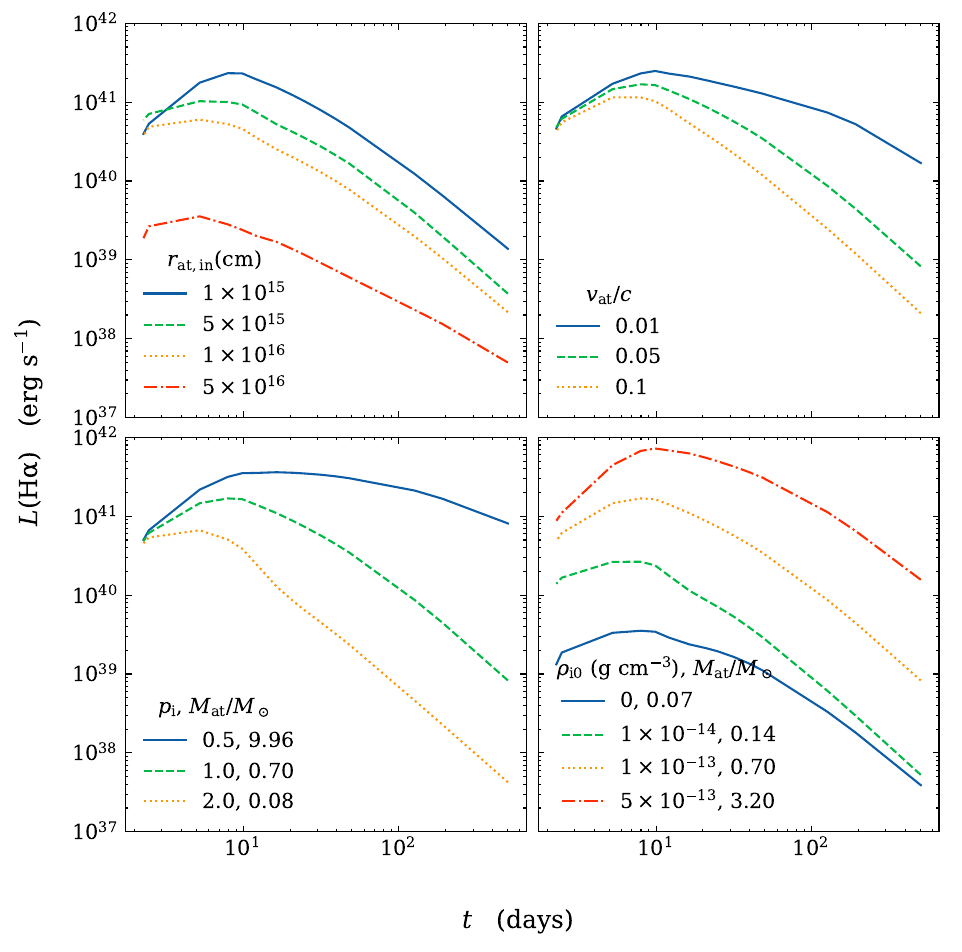}
\caption{Evolution of $\rm H\alpha$ luminosities with time for different model parameters. Top left, top right, bottom left and bottom right panels represent the different inner radii, velocities, power-law slope of density distribution, and densities at inner radius for atmosphere. Except for the varying parameter, the other parameters are fixed, where $\rho_\mathrm{i0}=1\times 10^{-13}\rm g\ {cm}^{-3}$, $v_\mathrm{at}=0.05c$, $r_\mathrm{at,in}=2\times 10^{15} \mathrm{cm}$, $p_\mathrm{i}=1.0$, $M_\mathrm{BH}=5\times 10^6 M_\odot$, $s=0.1$, $r_{\mathrm{l}0}=1200 r_\mathrm{s}$ and $p_\mathrm{rl}=0.8$ are adopted in our model. When we explore the impact of $r_\mathrm{at,in}$ on the results, in order to ensure that it is the position of the atmosphere rather than the total mass that affects the calculation results, we set $\rho_\mathrm{i0}r_\mathrm{at,in}^2 =4\times 10^{17} \rm g\ {cm}^{-1}$ instead of $\rho_\mathrm{i0}$ to remain unchanged during the calculation.
}
\label{fig:4}
\end{figure*}

Based on \texttt{CLOUDY} calculations, the strength of different emission lines can be calculated at given ionization spectrum and density distribution in the atmosphere. In this work, we analyze three typical lines as observed TDEs ($\rm H\alpha$, $\rm H\beta$ and \ion{He}{2} $4685.64\rm \AA$). In Figure \ref{fig:4}, we present the evolution of the luminosity of $\rm H\alpha$ for different parameters. Inspecting the top-left panel, it can be found that the $\rm H\alpha$ luminosity decreases with increasing inner radius of atmosphere, which is mainly caused by the lower ionization parameter (e.g., $L/r^2$). In the top-right panel, we show the results with different outflow velocity of atmospheres, where the higher velocity will lead to lower density and more outflowing matter will be ejected far away (i.e., lower ionization parameter). In the bottom-left panel, we present the results for different values of $p_i$ (or the total atmospheric mass $M_{\rm at}$). It can be found that $p_i$ cannot be too large since the atmospheric mass should be too heavier (e.g., $M_{\rm at}\sim10M_{\odot}$ \ for $p_i=0.5$), which will predict very high line luminosity even several hundreds of days later. In the bottom-right panel, we show the results for different initial parameters of $\rho_{\rm i0}$, which also regulate the mass of outer atmospheres, which are similar to that of different parameters of $p_i$. We find that our model can roughly reproduce the $\rm H\alpha$ line luminosities (e.g., peak $L_{\rm H \alpha}\sim$ several$\times 10^{41}\rm erg/s$) with reasonable model parameters. It should be noted that an atmosphere far from the photosphere or less mass in atmosphere will lead to a weaker intensity of lines. We also show one case for the atmosphere from pure disk winds (solid line in the bottom-right panel of Figure 4), where $L_{\rm H\alpha}< \rm several \times 10^{39}\rm erg\ s^{-1}$ and it is normally lower than the peak observational value by roughly two orders of magnitude. 

\begin{figure*}[ht!]
\plotone{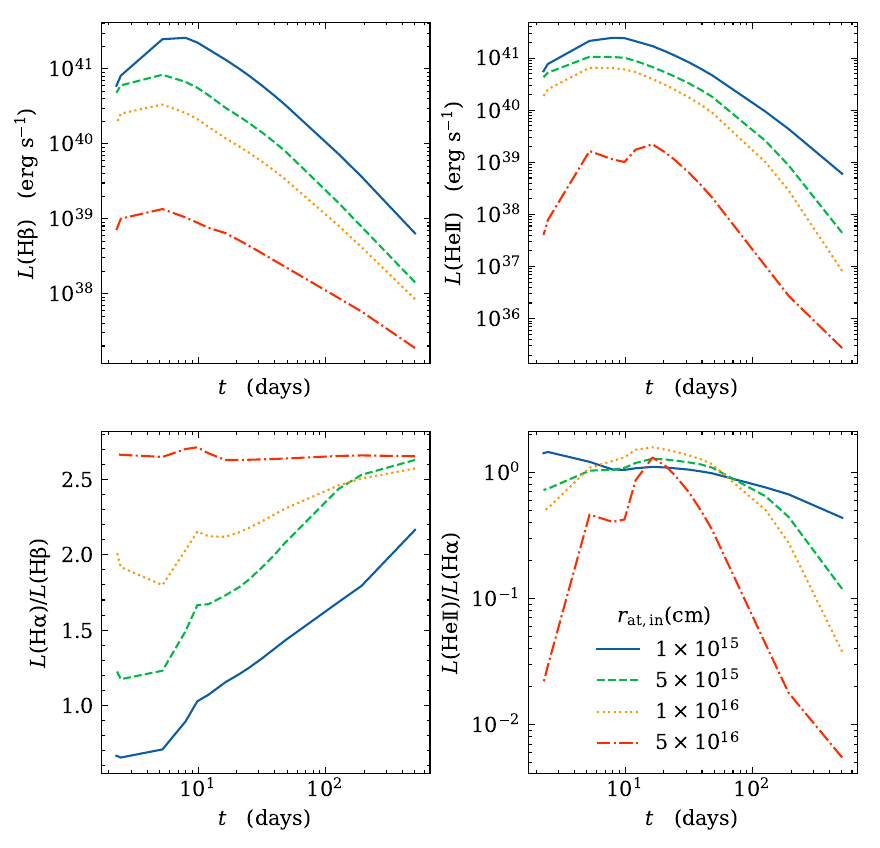}
\caption{Similar to Figure 4, but for the evolution of $\rm H\beta$ luminosities $L(\rm H\beta)$ (top-left panel), \ion{He}{2} $4685.64\AA$ luminosities $L(\rm He \ II)$ (top-right panel), $L(\rm H\alpha)/L(\rm H\beta)$ line ratios (bottom-left panel) and $L(\rm He \ II)/L(\rm H\alpha)$ line ratios (bottom-right panel) with time. In the calculation, we adopt $\rho_\mathrm{i0}r_\mathrm{at,in}^2 =4\times 10^{17} \rm g\ {cm}^{-1}$, $v_\mathrm{at}=0.05c$ and $p_\mathrm{i}=1.0$ respectively. 
}
\label{fig:5}
\end{figure*}

We also present the evolution of the $\rm H\beta$ luminosities, the \ion{He}{2} luminosities, the ratios $L(\rm H\alpha$)/$L(\rm H\beta)$ and $L(\rm He \ II)$/$L(\rm H\alpha)$ in Figure \ref{fig:5}. It can be found that the evolution of $L(\rm H\beta)$ and $L(\rm He \ II)$ is roughly similar to that of $L(\rm H\alpha)$ (top-left and top-right panels). In the bottom-left panel, the ratio $L(\rm H\alpha$)/$L(\rm H\beta)$ increases with time (or decreases with the accretion / outflow rates) for $r_{\rm at,in}<1\times 10^{16}\rm \ cm$, which is roughly unchanged as $L(\rm H\alpha$)/$L(\rm H\beta)\sim 2.7$ if $r_{\rm at,in}=5\times 10^{16}\rm \ cm$.  The ratio $L(\rm He \ II)$/$L(\rm H\alpha)$ roughly decays with time except for the case of $r_{\rm at,in}=5\times 10^{16}$ cm, where the very large radius of atmosphere may not happen in observed TDEs because of the quite low line luminosities.

\subsection{Comparison with AT2018dyb}

We compare our model with a TDE candidate AT2018dyb, which was discovered by the All-Sky Automated Survey for Supernovae \citep[ASAS-SN,][]{shappee2014man} in July 2018. Blackbody fitting shows that AT2018dyb has a peak luminosity of $L_\mathrm{p}\sim 10^{44} \ \mathrm{erg} \ \mathrm{s^{-1}}$, followed with a drop by about an order of magnitude within 100 days\citep{leloudas2019spectral}. This event shows strong emission lines, including hydrogen, helium, oxygen, and nitrogen lines\citep[see][for more details]{leloudas2019spectral}. The BH mass is $M_\mathrm{BH}=(1.3-8.3)\times 10^6 M_\odot$ based on the $M_{\rm BH}-\sigma_{*}$ relation, where $\sigma_{*}$ is the stellar velocity dispersion \citep[also see][for the details]{leloudas2019spectral}.

\begin{figure*}[ht!]
\plotone{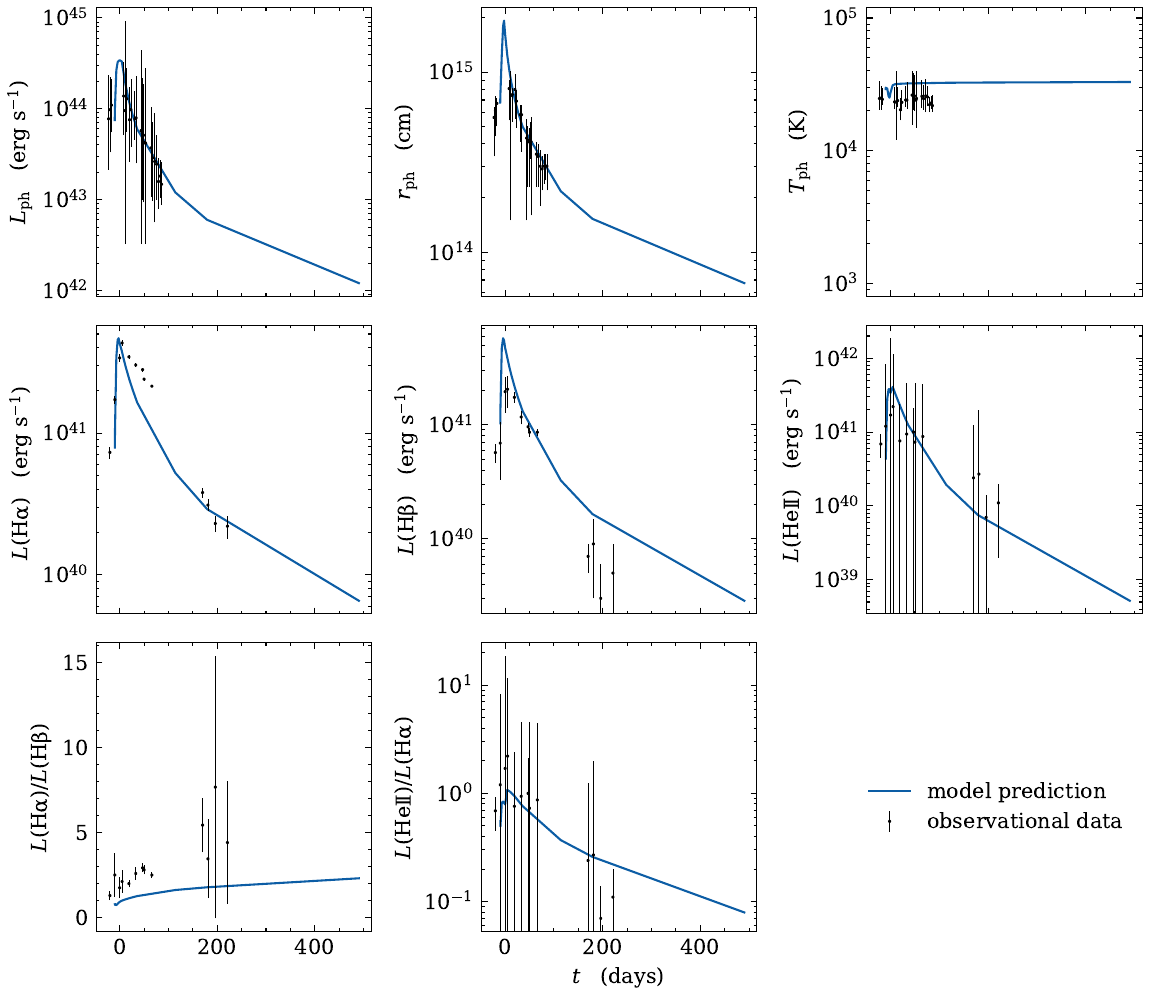}
\caption{Modelling results for AT2018dyb. The evolution of photosphere luminosity, radius, and temperature are shown in top panels. The line luminosities of $\rm H\alpha$, $\rm H\beta$, \ion{He}{2} $4685.64\AA$ are presented in the middle panels. In the bottom panels, the line ratios of $L(\rm H\alpha)/L(\rm H\beta)$ and $L(\rm He \ II)/L(\rm H\alpha)$ are given. The parameters of the accretion disk and wind are $M_\mathrm{BH}=5\times 10^6 M_\odot$, $s=0.15$, $r_{\mathrm{l}0}=1350 r_\mathrm{s}$ and $p_\mathrm{rl}=0.85$. In the atmosphere model, $v_\mathrm{at}= 0.05c$, $\rho_\mathrm{i0}=3\times 10^{-13}$ $\mathrm{g}$ $\mathrm {cm}^{-3}$, $p_\mathrm{i}=1$ and $r_\mathrm{at,in}=1.9\times 10^{15}\mathrm{cm}$ are adopted. The observational data in the figure are adopted from \citet{leloudas2019spectral} and \citet{charalampopoulos2022detailed}.}
\label{fig:6}
\end{figure*}

Based on observations of \cite{leloudas2019spectral} and \cite{charalampopoulos2022detailed}, we fit the observed properties of the photosphere, as shown in the top panels of Figure \ref{fig:6}. The parameters of $M_\mathrm{BH}=5\times 10^6 M_\odot$, $s=0.15$, $r_{\mathrm{l}0}=1350 r_\mathrm{s}$ and $p_\mathrm{rl}=0.85$ are adopted for the slim disk and wind-envelope model. With these parameters, our model can roughly reproduce the observational results. In the middle panels of Figure \ref{fig:6}, the luminosities of H$\alpha$, H$\beta$ and \ion{He}{2} are shown, while the line ratios of H$\alpha$/H$\beta$ and \ion{He}{2}/H$\alpha$ are presented in bottom panels. In the atmosphere model, $v_\mathrm{at}= 0.05c$, $\rho_\mathrm{i0}=3\times 10^{-13}$ $\mathrm{g}$ $\mathrm {cm}^{-3}$, $p_\mathrm{i}=1$ and $r_\mathrm{at,in}=1.9\times 10^{15}\mathrm{cm}$ are adopted. Based on the fitted spectrum of photosphere, we find that our model can roughly reproduce the evolution of several typical emission lines of AT2018dyb. It should be noted that the theoretical prediction of H$\alpha$/H$\beta\sim 1-3$, which is a little bit lower than that observational results (H$\alpha$/H$\beta\sim 1-5$).

\section{Conclusion and Discussion} \label{sec:4}

In this work, we study the properties of the envelope and emission lines for TDEs based on the global solutions of slim disk, where the mass loss rate due to the radiation pressure is calculated for the case of the super-Eddington accretion regime. The optical thick disk winds will absorb the disk radiation and reradiate it at optical wavebands, which naturally form the photosphere. We further calculate the line luminosities from the disk winds based on the \texttt{CLOUDY} modeling, and we find that the line luminosities are always lower than those of observations. We propose that the atmosphere outside the photosphere may be mainly due to the stellar debris collision. The envelope, with a typical mass of several tens percent of the solar mass, can roughly reproduce the observed line luminosities(e.g., $\rm H\alpha$, $\rm H\beta$ and \ion{He}{2}).

In this work, we adopt an analytical global solution for the slim disk to explore the properties of photosphere and line properties, where the disk-winds will be naturally launched if the accretion rate is larger than a critical value. With the decrease of the accretion rate, the wind will become weaker and weaker and, therefore, the envelope formed from the disk winds will shrink, which can roughly explain the decrease of the photosphere size as the decay of luminosity. According to \citet{cao2015limits} and \citet{feng2019global}, the wind will disappear when the accretion rate is lower than $\sim 1.7-1.9 \dot{M}_{\rm Edd}$ due to the transition to a thin accretion disk. However, the optical and X-ray observations suggested that the wind should still exist, and, therefore, we simply include the wind based on a power-law accretion rate. This will also contribute some matter to the envelope even in sub-Eddington regime. It should be noted that the global radiation MHD simulations will shed more light on the properties of the wind/outflow in super-Eddington accretion regime, where the wind strength, energy advection in disk, and the radiation spectrum may be different from the analytical solutions \citep[e.g.,][]{jiang2019super}. Both analytical work and simulation results suggest that an important fraction of gas feeding the accretion disk at outer radius is removed by the outflow. \citet{thomsen2022dynamical} conducted three-dimensional general relativistic radiation magnetohydrodynamic simulations of TDE accretion disks at varying accretion rates in the regime of super-Eddington accretion, where they found that the real accretion rate near the BH event horizon may be higher than the critical value as proposed by analytical work \citep[e.g.,][]{cao2015limits, feng2019global}. The strength of the disk wind/outflow will affect both the disk luminosity and the properties of the envelope, which will also affect the possible evolution of size, temperature and luminosity of the envelope (e.g., Figure 3). 
It should be pointed out that the simulation resolution is still limited and more high-resolution simulations are expected to further test this issue.

We explore the evolution of envelope and broad lines in TDEs based on the disk-wind model. We find that the evolution of the envelope from the optically thick wind is roughly consistent with the observational constraints, and, however, the luminosities of broad lines are much weaker compared to the observations if only the photoinozation of the winds outside of the envelope is considered. This conclusion is roughly consistent with former results, where either the atmosphere with very large size \citep[e.g., $>10^{17}$ cm,][]{parkinson2022optical} or strong wind from a very large disrupted star \citep[e.g.,][]{mageshwaran2023probing} is needed to produce the observed line luminosities. It has been widely believed that the interactions between the debris streams will lead to a self-crossing shock, which will dissipate the kinetic energy of the debris. The exchange of angular momentum and circularization during the debris collision will lead to the formation of an outflow and an accretion disk \citep[e.g.,][]{hayasaki2013finite, bonnerot2016disc, skadowski2016magnetohydrodynamical, liptai2019disc, lu2020self, bonnerot2021first}. 
In this work, we find that the peak luminosities of emission lines may be mainly produced from the atmosphere that formed during the earlier stream-stream collision. The time scale for debris circularization and disk formation is still unknown, where recent simulations show that a large fraction of debris material can form disks with a dynamical timescale \citep[e.g., a couple of days for typical TDE parameters, ][]{curd2021global, andalman2022tidal}. If we simply adopt 10 days as an approximation, the atmosphere formed from the debris outflow can reach $\sim 10^{15}$ cm for an outflow velocity of $0.05 c$ \citep[e.g.,][]{charalampopoulos2022detailed}, which normally stay outside of the envelope. We neglect the debris-collision material to form the observed TDE envelope, where we assume the material with low and high angular momentum during the debris collision forms the accretion disk and the outflow separately. A simple power-law density profile (e.g., $\rho \propto r^{-2}$) of the atmosphere from steady-state wind is widely adopted in previous works. However, the outflow rate is strongly decaying with time, the density distribution along the radius should not follow a simple power-law. The expanding matter from the highest outflow rate may dominate the contribution to the line emission, which also depends on the expanding velocity or distance from the central engine. This is the difference from our results to former works. It should be noted that the atmosphere mass and the line luminosities are depend on multiple parameters of $p_i$,  $v_{\rm at}$ and $\rho_{\rm i0}$, where the line luminosities are sensitive to the gas distribution or incident fluxes of ionizing photons (e.g., $1/r^2$). The atmosphere gas with higher density and closer to the photosphere will produce stronger line luminosities. For example, we find that a smaller parameter of $p_i=-1$ with a higher value of $\rho_{\rm i0}=6\times 10^{-3}\rm g\ cm^{-3}$ and slower outflow velocity $v_{\rm at}=0.01c$ will lead to similar line luminosities (e.g., peak luminosity of $L(\rm H \alpha)\sim$ several$\times 10^{41}\rm erg/s$ as shown in Figure \ref{fig:4}), where the atmosphsere mass $M_{\rm at}\sim 0.3 M_{\odot}$. More observations and model fittings are requested to further test our model (e.g., break the possible degeneracies in model parameters), which will be our future work.

The formation and geometry of the atmosphere outside the photosphere are unclear. In some of the previous works, the emission lines are calculated from the disk-wind and envelope model based on \texttt{CLOUDY} photoionization\citep[e.g.,][]{mageshwaran2023probing}. The calculation of photoionization and radiative transfer in the optically thick envelope is still difficult to be done, and the results may not be very reasonable because the \texttt{CLOUDY} code is mainly developed for optically thin clouds even though the optical depth effect is considered \citep{roth2016x, ferland20172017}. In our model, we do not consider the line emission from the optically thick envelope and mainly consider the debris-collision-induced atmosphere outside of the photosphere. We find that the atmosphere is normally optically thin with an effective optical depth of less than a few percent. However, the electron scattering optical depth in atmosphere is up to several in most cases, and it can reach 10 or even 15 near the peak luminosities ($t\sim$10 days), where too large values may predict less accurate line emission near the peak luminosities (several days). In the bottom-left panel of Figure \ref{fig:5}, the ratio $L(\rm H\alpha)/L(\rm H\beta)$ increases with time (or weaker winds) except for $r_{\rm at,in}=5\times 10^{16}$ cm. The ratio $L(\rm H\alpha)/L(\rm H\beta)$ might be affected by high density effects or wavelength-dependent extinction by the atmosphere \citep[see also this effect in AGNs, e.g.,][]{osterbrock1984active, goodrich1995dust, gaskell2017case}. \citet{pottasch1960balmer} have suggested that higher optical depths in the Balmer lines would lead to steeper $L(\rm H\alpha)/L(\rm H\beta)$ ratios. This is because when the lower Lyman lines have higher optical depths, $\rm H\alpha$ is scattered but H$\beta$ can be converted to P$\alpha$ and H$\alpha$, which will reduce the number of H$\beta$ photons and increases the number of H$\alpha$ photons. The atmosphere at a larger radius has lower density and temperature, which normally predicts a higher value of the $L(\rm H\alpha)/L(\rm H\beta)$ ratio ($\sim 2.7$), roughly corresponding to the Case B value of 2.74 \citep{osterbrock2006astrophysics}. The $L(\rm He \ II)/L(\rm H\alpha)$ ratio is normally positively correlated with the photosphere luminosity(or decay with time, see bottom-right panel of Figure \ref{fig:5}). The radius of the Stromgren sphere of \ion{He}{2} is larger than or comparable to the thickness of the atmosphere, so the emission of \ion{He}{2} is comparable to H$\alpha$ at the beginning with a smaller size of atmosphere. Although the size of atmosphere increases with time, the Stromgren sphere of H is significantly larger than that of He, leading to a lower $L(\rm He \ II)/L(\rm H\alpha)$ ratio(see Figures 5 and 6). \citet{roth2016x} also proposed that the strength of emission lines and the line ratios in TDEs will be set by the wavelength-dependent optical depths, where the hydrogen Balmer line emission is often strongly suppressed relative to the helium line emission at higher optical depth. We note that our model is still simple, where the disk wind model in the phase of a very high Eddington rate and the evolution of parameters along time are not well understood. Further work based on simulations can provide insight into the formation and evolution of the TDE emission lines, which will be our future work.

\begin{acknowledgments}
We thank the referee for constructive suggestions that help us to improve our results. The work is supported by the National Natural Science Foundation of China (grants 12233007 and U1931203) and the science research grants from the China Manned Space Project (No. CMS-CSST-2021-A06). The authors acknowledge Beijng PARATERA Tech CO., Ltd. for providing HPC resources that have contributed to the results reported within this paper.
\end{acknowledgments}

\bibliography{main}{}
\bibliographystyle{aasjournal}

\end{document}